\documentclass[conference]{IEEEtran}
\IEEEoverridecommandlockouts
\usepackage{cite}

\usepackage{layouts}

\usepackage{cite}
\usepackage{amsmath,amssymb,amsfonts}
\usepackage{amsthm}
\usepackage{algorithmic}
\usepackage{graphicx}
\graphicspath{{.},{pdf/}}
\DeclareGraphicsExtensions{.pdf,.jpeg,.png}
\usepackage{textcomp}
\usepackage{xcolor}
\usepackage{algorithm}
\usepackage{url}
\usepackage{tikz}

\newtheorem{alg}{Algorithm}
\newtheorem{condition}{Condition}

\usepackage{xspace}

\newcommand{\pdtau}[1][d,\tau]{\ensuremath{\mathbf{p}_{#1}}\xspace}

\newcommand{\R}{\ensuremath{\mathbf{R}}\xspace}
\newcommand{\curve}{\ensuremath{\mathcal C}\xspace}
\newcommand{\cpdtau}{\ensuremath{\curve_{p,d,\tau}}\xspace}
\newcommand{\Qpdtau}{\ensuremath{Q_{p,d,\tau}}\xspace}

\newcommand{\est}[1]{\ensuremath{\hat{#1}}\xspace}
\newcommand{\estrc}{\ensuremath{\est{\rv}_\curve}\xspace}
\newcommand{\vect}[1]{\ensuremath{\mathbf{#1}}\xspace}

\newcommand{\pv}{\vect{p}}

\newcommand{\qv}{\vect{q}}

\newcommand{\rv}{\vect{r}}
\newcommand{\dd}{\ensuremath{\mathrm{d}}}

\newcommand\copyrighttext{%
  \footnotesize \textcopyright 2023 IEEE.\@ Personal use of this material is permitted.
  Permission from IEEE must be obtained for all other uses, in any current or future
  media, including reprinting/republishing this material for advertising or promotional
  purposes, creating new collective works, for resale or redistribution to servers or
  lists, or reuse of any copyrighted component of this work in other works.%
}
\newcommand\copyrightnotice{%
\begin{tikzpicture}[remember picture,overlay]
\node[anchor=south,yshift=10pt] at (current page.south) {\fbox{\parbox{\dimexpr\textwidth-\fboxsep-\fboxrule\relax}{\copyrighttext}}};
\end{tikzpicture}%
}
\begin{document}
%
\title{Statistical reconstruction of pulse shapes\\ from pulse streams%
}

\author{\IEEEauthorblockN{Marek W. Rupniewski}
    \IEEEauthorblockA{%
        \textit{Institute of Electronic Systems}\\
        \textit{Warsaw University of Technology}\\
        Nowowiejska 15/19, 00-665 Warsaw, Poland\\
        Email: Marek.Rupniewski@pw.edu.pl\\%
        ORCID: 0000-0003-3861-510X%
}}%

\maketitle
\copyrightnotice

\begin{abstract}
    A short sample sequence of a finite-length pulse signal allows for its reconstruction only if the signal has a sparse representation in some basis. The recurrence of the pulse allows for a statistical approach to its reconstruction. We propose a novel method for this task. It is based on the distribution of short sample sequences treated as points that lie along a closed curve in a low-dimensional Euclidean space. We prove that the probability distribution of the points along this curve determines the underlying pulse signal uniquely. Based on this discovery, we propose an algorithm for pulse estimation from a finite number of short sequences of pulse-stream samples.

\end{abstract}


\section{Introduction}
Signals comprised of a stream of short pulses, referred to as pulse streams,
appear in many applications, including biomedicine~\cite{Rudresh18},
radar~\cite{Pan17}, and ultrasonics~\cite{Tur11,Saurav20}. 
In general, Shannon sampling theorem requires a high sampling rate 
for the good-quality reconstruction of short pulses. If the
pulses have a sparse representation in some basis, then it is possible to
reconstruct them from samples taken at a much lower sub-Nyquist frequency by 
the compressed sampling technique \cite{Tur11,Matusiak12,Hedge11}.
Sub-Nyquist sampling is also possible if the pulse shapes are known up to their
amplitudes and inter-pulse distances \cite{Meisam20,Huang22}. In our study, we
do not assume prior knowledge of the pulse shape nor restrict the pulse
to have a sparse representation. Instead, we assume that the pulse stream
consists of a pulse that recurs in time, and we set the reconstruction of the
shape of this single pulse as our goal. The main contribution of this paper is as follows:
\begin{itemize}
    \item It shows that a single pulse can be reconstructed from the
        probability distribution of short sample sequences taken at a~sub-Nyquist
        rate.
    \item It shows how to estimate the pulse from a finite number
        of these sequences.
\end{itemize}
We propose reconstruction algorithms based on the
study of the probability distribution of short sample sequences, referred to as
sample trains. Such an approach was introduced in our papers~\cite{Rup20icassp,Rup21icassp,Rup21spl,Rup21iet} in the context of periodic signals.

The paper is organized as follows. The next section shows the relationship
between sample trains, pulse signals, and pulse streams.
Sections~\ref{sec:distribution} and \ref{sec:reconstruction} are devoted to
the algorithms for pulse reconstruction from the probability distribution of
sample trains and a finite number of sample trains, respectively. Section~\ref{sec:simulation} presents the results of numerical simulations of the latter algorithm. 
The paper is concluded in Section~\ref{sec:conclusion}.
\IEEEpubidadjcol

\section{Pulse signals and pulse streams}\label{s:signals}
\begin{figure*}[tp]
    \centering
    \includegraphics[]{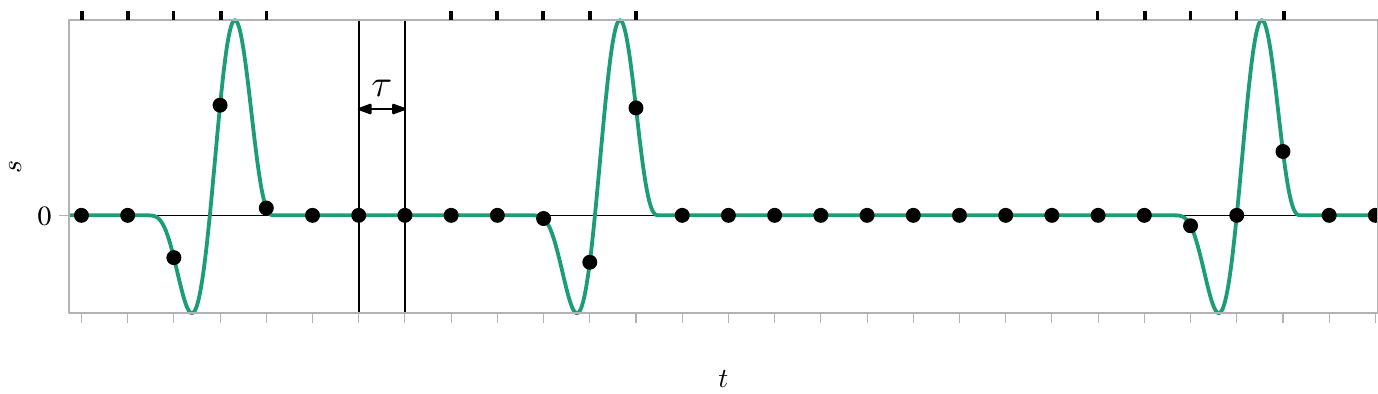}
    \includegraphics[]{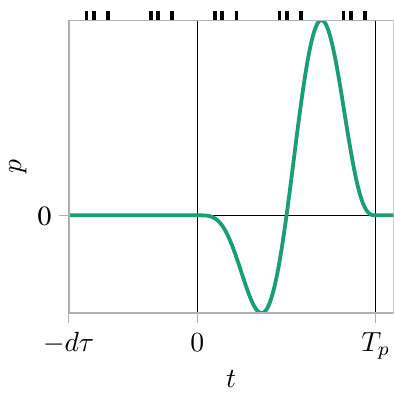}
    \vspace{-2em}
    \caption{A pulse stream signal $s$ (to the left) and the underlying pulse signal $p$ (to the right).
    Small circles indicate the samples of the pulse stream.
    Ticks along the top edges of the plots show the corresponding starting times of non-zero
    sample trains of length~$3$ and time-span $2\tau$.}%
    \label{fig:pulses}\label{fig:pulse1}
    \vspace{2em}
    \includegraphics[]{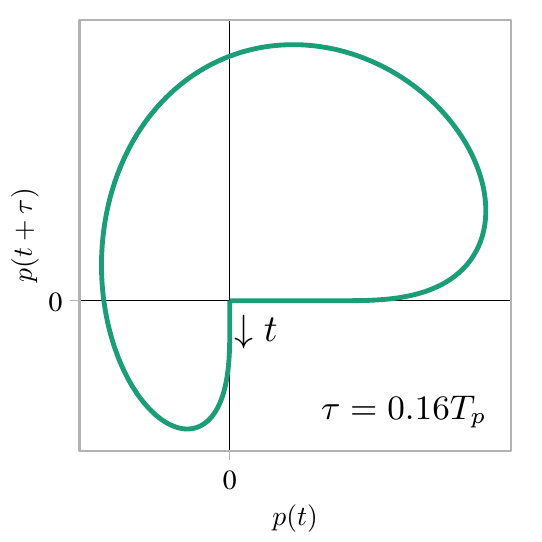}
    \includegraphics[]{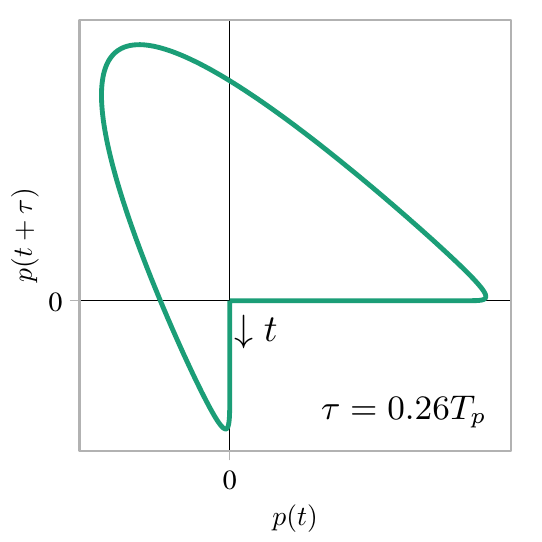}
    \includegraphics[]{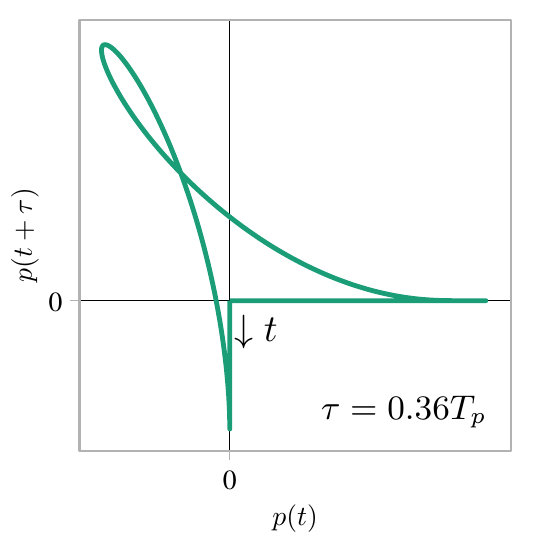}
    \vspace{-1em}
    \caption{The images of \pdtau[2,\tau] for signal $p$ shown in Fig.~\ref{fig:pulse1}, and three inter-sample distances $\tau = 0.16 T_p$, $\tau = 0.26 T_p$, $\tau=0.36 T_p$}
    \label{fig:curves}
\end{figure*}

We define pulse signals as continuous-time signals with finite support.
Without loss of generality, we assume that for any
pulse signal $p$, the smallest interval outside of which the signal
vanishes is of the form $(0,T_p)$, where $T_p > 0$ is the pulse length; see Fig.~\ref{fig:pulse1}.
A pulse stream is a signal that consists of copies of the same pulse signal,
i.e., it is a signal of the following form, where $N$ may take a finite or $+\infty$ value.
\begin{equation} \label{e:repetitive}
    s(t) = \sum_{i=1}^N p(t-t_i),\qquad t_1 < t_2 < \dots.
\end{equation}
Fig.~\ref{fig:pulses} shows an example of a pulse stream.
The minimum distance between pulses that form signal $s$ is termed inter-pulse distance and is denoted by $\Delta_s$, i.e.,
\begin{equation} \label{e:ipd}
    \Delta_s = \min_{i=1,\dots,N} (t_{i+1} - t_i - T_p).
\end{equation}

Sample trains of signal $p$ are defined as vectors of the form
\begin{equation}\label{e:pdtau}
    \pdtau(t) = [p(t),\,p(t+\tau),\,p(t+2\tau),\,\dots,\,p \left( t+d\tau \right)],
\end{equation}
where $d\tau$ is the time span of the sample train, $t$ is its starting time,
and $\tau$ is the inter-sample time distance. Note that sample trains \pdtau consist of $d+1$ samples. Therefore, $\pdtau(t)$ can be
treated as points in a $d+1$-dimensional Euclidean space.
Assume that a pulse signal $p$ recurs in time, in a periodic or non-periodic manner,
forming a pulse stream $s$. By
sampling $s$ uniformly, one may acquire multiple sample trains of
signal~$p$ which correspond to various starting times~$t_0\in\left(-d\tau,
T_p\right)$; see Fig.~\ref{fig:pulses}. 
In particular, the sets of sample
trains for the pulse stream and the underlying pulse signal coincide, provided that the following condition is satisfied.
\begin{condition}\label{cond:delta}
    $
        \Delta_s\geq d\tau.$ 
\end{condition}

\section{Distribution of sample trains}\label{sec:distribution}
If $p$ is a pulse signal of length $T_p$, then $\pdtau(t)$ is a train of zeros
unless $-d\tau<t<T_p$. When $t$ varies in the range $(-d\tau,T_p)$, then
$\pdtau(t)$ moves along a curve that lies in $\R^{d+1}$. Henceforth, this curve
is denoted by \cpdtau. Three such curves are shown in Fig.~\ref{fig:curves}.

A continuous mapping $\qv\colon(0,a)\to\cpdtau$, $a>0$, will be called a regular parameterization of curve \cpdtau if $\qv$ is a one-to-one function onto \cpdtau.
In this study, we focus on pulse signals $p$, inter-sample distances $\tau$, and
sample-train lengths $d+1$ that satisfy the following condition.
\begin{condition}\label{cond:reg}
    \pdtau restricted to interval $(-d\tau, T_p)$ is a regular parameterization of curve \cpdtau.
\end{condition}
The main obstacle to Condition~\ref{cond:reg} is the presence of
self-intersections of curve~\cpdtau. For example, signal from
Fig.~\ref{fig:pulse1} satisfies Condition~\ref{cond:reg}
for $d=2$ and $\tau=0.16T_p$ and it does not for $\tau=0.36T_p$ (\cpdtau
has self-intersections in the latter case as shown in the last graph
of Fig.~\ref{fig:curves}).
If Condition~\ref{cond:reg} is satisfied, then there exist exactly two arc-length parameterizations of curve $\curve = \cpdtau$. They have opposite orientations.
The initial part of curve \cpdtau can be tell from its terminal part by noting that the part that corresponds to $t < -d\tau + \tau$ lies in the $X_{d+1}$-axis, and the
part that corresponds to $t>T_p-\tau$ lies in the $X_1$-axis; see
Fig.~\ref{fig:curves}. Let $\rv_{\curve}\colon(0,L)\to\R^{d+1}$ be the unique arc-length parameterization of \curve that maps initial segments of
interval $(0,L)$ to initial segments of \curve ($L$ is the total length of~\curve).
The probability distribution of starting times $t$ of sample trains $\pdtau(t)$ results in the corresponding probability distribution of the trains treated as points of curve \cpdtau. 
By using parameterization $\rv_\curve$ we may pull-back this probability
distribution from curve \curve to interval $(0,L)$. In particular, the uniform
distribution of starting times $t\in(-d\tau,\,T_p)$ determines a quantile function $\Qpdtau\colon[0,1]\to[0,L]$. 
Quantile of order $\alpha$ of this distribution corresponds to starting time $t$ that lies in the $\alpha$ part of the distance from the left towards the right end of interval $(-d\tau,\,T_p)$, i.e.,
\begin{equation}\label{e:rvdef}
    \rv_\curve\left(\Qpdtau(\alpha)\right) = \pdtau\left(-d\tau + \alpha(T_p + d\tau)\right).
\end{equation}
Let $\pi_k$ denote a function that picks the $k$-th coordinate of its argument, i.e., 
$\pi_k\left([x_1,\dots,x_m]\right)=x_k$.
The $k$-th coordinate of function $\pdtau$, i.e., $\pi_k(\pdtau)$ vanishes outside the interval
\begin{equation}\label{e:interavalIk}
    I_k = \bigl(\left(1-k\right)\tau,\, T_p+\left(1-k\right)\tau\bigr)
.\end{equation}
Let $\alpha_{k,\min}$ and $\alpha_{k,\max}$ denote the fractions of interval $(-d\tau,\,T_p)$ that correspond to the ends of $I_k$, i.e.,
\begin{equation}\label{e:akmin}
    \alpha_{k,\min} = \frac{(d + 1-k)\tau}{T_p + d\tau},\,
    \alpha_{k,\max} = \frac{T_p + (d+1-k)\tau}{T_p + d\tau}.
\end{equation}
Consequently,
\begin{align}
    T_p &= \frac{(d+1-k)\tau}{\alpha_{k,\min}}-d\tau,& 
    k&=1,\dots,d,\label{e:Tp1}\\
    T_p &= \frac{(k-1)\tau}{1-\alpha_{k,\max}}-d\tau,&
    k&=2,\dots,d+1\label{e:Tp2}.
\end{align}
Let
\begin{equation}\label{e:qvdef}
    \qv(\alpha) = \begin{cases}
        \rv_\curve\left(\Qpdtau(\alpha)\right) & \alpha\in(0,1)\\
        0 & \alpha \geq 1.
    \end{cases}
\end{equation}
By~\eqref{e:rvdef},
   $\qv(\alpha) = \pdtau\left(-d\tau + \alpha(T_p + d\tau)\right)$. Therefore, by~\eqref{e:pdtau}
\begin{multline} \label{e:ptalter}
    p(t) = 
    \pi_k \bigl( \pdtau \left( t - (k-1)\tau \right)\bigr) \\
    = \pi_k \left( \qv\left(\frac{t + (d + 1 -k)\tau}{T_p + d\tau}\right)\right),
    k=1,\,\dots,\,d + 1.
\end{multline}
Eqs.~\eqref{e:akmin}--\eqref{e:ptalter} result in
the following reconstruction algorithm.
\begin{alg}\label{alg:fromdensity}
    \par {\bf{Inputs:}} Sample period $\tau$ and the probability distribution (supported on a~curve~\curve) of sample trains $\pdtau(t)$ that result from the uniform distribution of starting times $t$ for a pulse signal $p$ that satisfies Condition~\ref{cond:reg}.
    {\bf{Output:}} pulse signal $p$.
    \begin{enumerate}
        \item Take the arc-length parameterization
            $\rv_{\curve}\colon(0,L)\to\curve$ that maps the initial segments of
            interval $(0,L)$ to the initial segments of \curve.  
        \item Compute the quantile function $Q\colon[0,1]\to[0,L]$ of the $\rv_{\curve}$-pull-back of the given probability distribution.
        \item Define mapping $\qv$ according to~\eqref{e:qvdef} (by the assumption $\Qpdtau=Q$) and find $T_p$
            by any of Eqs.~\eqref{e:Tp1} or \eqref{e:Tp2}.
        \item Compute signal $p$ by Eq.~\eqref{e:ptalter} for any $k=1,\dots,d+1$.
    \end{enumerate}
\end{alg}

\section{Pulse reconstruction from sample trains}\label{sec:reconstruction}
The previous section shows how to reconstruct a pulse signal $p$ from the probability distribution supported on curve $\curve = \cpdtau$.
In practice, instead of such distribution and curve, we can have only a finite
number of non-zero sample trains:
\begin{equation}
    \label{e:pn}
    \pv_1,\,\pv_2,\,\dots,\,\pv_n\in\R^{d+1}.
\end{equation}
In this section, we show how to construct a reliable
estimator of the pulse signal in such a practical scenario. We will approach the goal in three steps. First, we explain how to estimate curve \cpdtau.
Then, we show how to estimate the probability distribution on this curve. Eventually, we will adjust~Algorithm~\ref{alg:fromdensity} to accomplish the goal.
\subsection{Curve estimation}\label{ss:curvest}
We can approximate curve \cpdtau with a polygonal chain by connecting
points~\eqref{e:pn} with line segments. To get a reasonable
approximation, we need to find the order in which these points lie on
curve~\cpdtau. This can be accomplished, e.g., with the NN-CRUST
algorithm~\cite{dey2000} or its improved version presented in~\cite{lenz05}.
The NN-CRUST algorithm is proven to give the correct order of points along a curve,
provided that the curve is sampled densely enough. However, it does not
identify the endpoints of curve \cpdtau. Therefore, we complement the algorithm
with the following postprocessing. Assume that the result of the NN-CRUST algorithm
is an ordering of points~\eqref{e:pn} given by permutation $\eta$, i.e.,
$\pv_{\eta_1},\,\dots,\,\pv_{\eta_n}$.
Let $\pv_{\eta_k}$ be the point that lies on $x_{d+1}$-axis in the closest distance
to the origin of the coordinate system. If $\pv_{\eta_{k+1}}$ lies on
$x_{d+1}$-axis as well, or if $\pv_{\eta_{k-1}}$ lies on $x_1$-axis, then we
replace permutation $\eta$ with $\eta'$ defined as
\begin{equation}
    \label{e:etap1}
    \eta'_1 = \eta_k,\, \eta'_2 = \eta_{k+1},\,\dots,\eta'_n=\eta_{k+n-1},
\end{equation}
where arithmetic operations on the indices are
performed modulo~$n$, e.g., $\eta_{n+1}=\eta_1$.
In the opposite case, i.e., if neither $\pv_{\eta_{k+1}}$ lies on $x_{d+1}$-axis nor $\pv_{\eta_{k-1}}$ lies on $x_1$-axis, we set
\begin{equation}
    \label{e:etap2}
    \eta'_1 = \eta_k,\, \eta'_2 = \eta_{k-1},\,\dots,\eta'_n=\eta_{k-n+1}.
\end{equation}
If NN-CRUST does not identify points~\eqref{e:pn} as belonging to a single
closed curve or if no point of~\eqref{e:pn} lies on $x_d$-axis then we cannot determine the order in which points~\eqref{e:pn} lie on curve~\curve and we have to stop the reconstruction procedure until more points are available.

\subsection{Probability distribution estimation}\label{ss:distest}
Points $\pv_{\eta'_1},\,\dots,\,\pv_{\eta'_n}$ obtained in the previous
subsection can be used as vertices of a polygonal chain that starts and ends
at the origin. This chain  approximates curve \cpdtau. Let
$\estrc\colon(0,\est{L})\to\R^{d+1}$ be the arc-length parameterization of
this chain. In particular,
\begin{equation}
    \label{e:estL}
        \est{L} = \|\pv_{\eta'_1} - 0\| + 
        \sum_{k=1}^n\|\pv_{\eta'_{k+1}} - \pv_{\eta'_k}\|
        + \|0 - \pv_{\eta'_n}\|.
\end{equation}
We estimate the quantile function \Qpdtau introduced in
Section~\ref{sec:distribution} by first defining $\est{Q}$ on a grid:
$\est{Q}\left(0\right) = 0$, $\est{Q}\left(1\right) = \est{L}$, and
\begin{equation}
    \label{e:estQdef1}
    \est{Q}\left(\frac{k-0.5}{n}\right) = 
        \estrc^{-1}\left(\pv_{\eta'_k}\right),\, k=1,\,\dots,\,n.
\end{equation}
Then, we linearly interpolate $\est{Q}$ between the nodes of the grid.
Eventually, we obtain an approximation $\est{\qv}$ of mapping $\qv$
by~Eq.~\eqref{e:qvdef}, in which we replace, \Qpdtau and $\rv_\curve$ by $\est{Q}$ and \estrc, respectively.

\subsection{Pulse-reconstruction algorithm}
Let $n_{k,\min}$ and $n_{k,\max}$ be the ordinal numbers of, respectively, the first and the last non-zero entries among
\begin{equation}
    \pi_k \left( \pv_{\eta'_1}\right),\,\dots,\, \pi_k \left( \pv_{\eta'_n}\right).
\end{equation}

We define the following estimators
\begin{align}
    \est{\alpha}_{k,\min} &= \frac{n_{k,\min}-1}{n}, & k&=1,\,\dots,\,d,
    \label{e:akminest}\\
    \est{\alpha}_{k,\max} &= \frac{n_{k,\max}}{n}, & k&=2,\,\dots,\,d+1.
    \label{e:akmaxest}
\end{align}
and average Eqs.~\eqref{e:Tp1} and \eqref{e:Tp2} to get a $T_p$ estimator:
\begin{equation}
    \label{e:Tpestim}
    \est{T}_p = \frac{\tau}{2d} \left( \sum_{k=1}^{d} \frac{d+1-k}{\est{\alpha}_{k,\min}} + \sum_{k=2}^{d+1} \frac{k-1}{1-\est{\alpha}_{k,\max}} \right) -d\tau
\end{equation}
Eventually, we estimate the pulse signal by averaging~\eqref{e:ptalter}, i.e.,
\begin{equation}\label{e:estptrecon}
    \est{p}(t) = \frac{1}{d + 1}\sum_{k=1}^{d + 1}\pi_k \left( \est{\qv}\left((t+(d + 1 -k)\tau)/(\est{T}_p + d\tau)\right)\right).
\end{equation}

The following algorithm concludes this section.
\begin{alg}\label{alg:fromtrains}
    \par {\bf{Inputs:}} parameters $d$ and $\tau$, sample trains~\eqref{e:pn} of a pulse signal $p$ that
    satisfies Condition~\ref{cond:reg}.
    {\bf{Output:}} estimator $\est{p}$ of pulse signal $p$
    \begin{enumerate}
        \item Order points~\eqref{e:pn} by the NN-CRUST algorithm.
        \item Find permutation~$\eta'$ as explained in Subsection~\ref{ss:curvest}.
        \item Compute the arc-length parameterization $\estrc$ of the polygonal chain with vertices $0,\,\pv_{\eta'_1},\,\dots,\,\pv_{\eta'_n},\,0$.
        \item Compute quantile function $\est{Q}$ by~\eqref{e:estQdef1}.
        \item Compute function $\est{\qv}$ by~\eqref{e:qvdef} in which $\qv$,
            $\rv_\curve$, and \Qpdtau are replaced with \est{\qv}, \estrc, and
            \est{Q}, respectively.
        \item Compute pulse length estimator $\est{T}_p$ by~\eqref{e:Tpestim}.
        \item Compute estimator $\est{p}$ by~\eqref{e:estptrecon}.
    \end{enumerate}
\end{alg}
Let us highlight that in Algorithm~\ref{alg:fromtrains}, the sample trains may also result from sampling a pulse stream, provided that this pulse stream satisfies Condition~\ref{cond:delta}.

\section{Simulation}\label{sec:simulation}
\begin{figure}[tp]
    \includegraphics[]{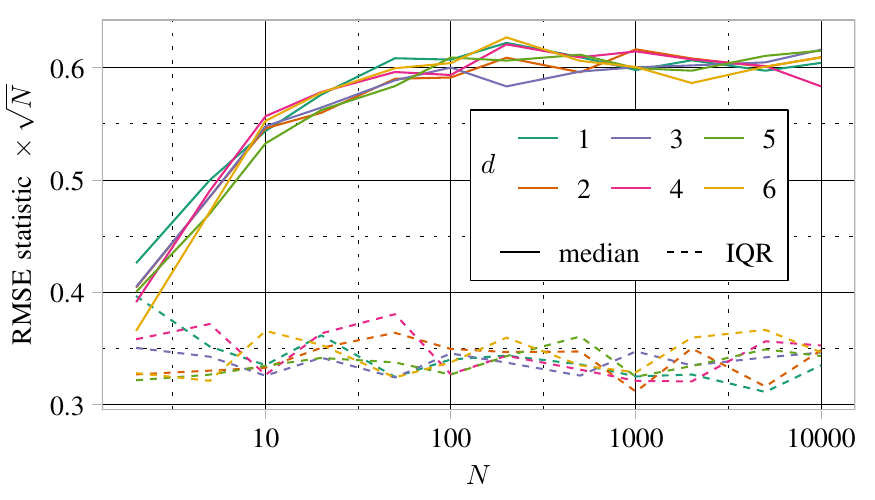}
    \caption{Median and inter-quartile range (IQR) of the RMSE of Algorithm~\ref{alg:fromtrains}}
    \label{fig:grnpul1}
\end{figure}

We have conducted a series of numerical simulations
to verify the reconstruction algorithm and assess its asymptotic behavior. We collected non-zero sample trains
of a pulse stream with the underlying pulse signal presented in Fig.~\ref{fig:pulse1}. For each of the considered values of sample train length $d+1$, and number of pulse recurrences $N$, we repeated the experiment $1000$ times to examine the statistical behavior of the root-mean-square reconstruction error defined as
\[
    \text{RMSE} = \left(\frac{1}{\max(T_p, \est{T}_p)}\int_0^{\max(T_p, \est{T}_p)} \left( p(t) - \est{p}(t) \right)^2 \dd t\right)^{1/2}
.\] 
As expected, the bigger the number of pulses $N$ is, the smaller the error is,
as the accuracy of the polygonal-chain approximation and the quantile-function
estimation rise with $N$.
Fig.~\ref{fig:grnpul1} shows that the median and inter-quartile range of the RMSE converge asymptotically to zero at the rate of the order $N^{-\frac{1}{2}}$.

\begin{figure}[tp]
    \centering
    \includegraphics[]{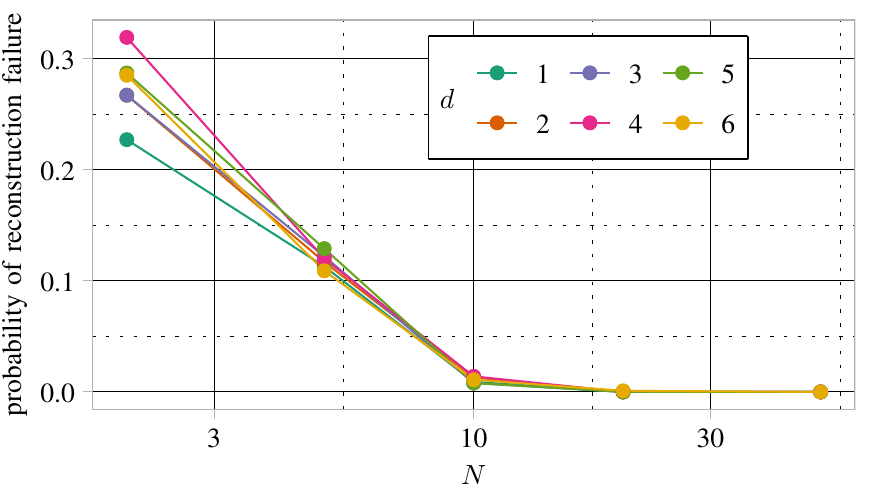}
    \caption{Probability of Algorithm~\ref{alg:fromtrains} stop due to insufficient data}
    \label{fig:prob1}
    \vspace{2em}
    \includegraphics[]{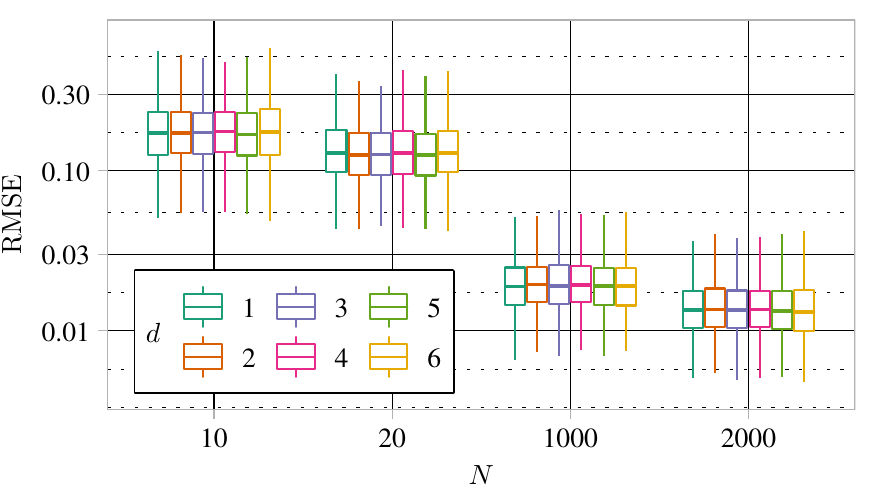}
    \caption{Boxplot of the RMSE of Algorithm~\ref{alg:fromtrains}}
    \label{fig:boxplot1}
\end{figure}

In general, the length of the trains $d$ cannot be too small. Otherwise, curve \curve
may fail to satisfy Condition~\ref{cond:reg}. 
On the other hand, too big $d$ can lead to violation of
Condition~\ref{cond:delta}. Moreover,
the bigger $d$ is, the longer curve \curve is, which is unfavorable for the
NN-CRUST algorithm and for the curve approximation with a polygonal chain.
Consequently, the probability of the reconstruction algorithm failure is bigger
for larger $d$; see Fig.~\ref{fig:prob1}. Fortunately, this probability drops
very quickly with $N$. Fig.~\ref{fig:boxplot1} shows no significant
differences between performance of the reconstruction algorithm for various
values of~$d$ when $N \gg 10$.
A quantitative analysis on the role of all possible factors on the quality of
the reconstruction is beyond the scope of the his study. However, we must
note that the RMSE also depends on the inter-sample distance $\tau$, which
affects the length of curve~\curve and may cause a violation of
Conditions~\ref{cond:delta} and~\ref{cond:reg} (see Fig.\ref{fig:curves}).
Eventually, the performance of the algorithm depends on the particular pulse
signal $p$ itself.


\section{Conclusion}\label{sec:conclusion}
We have proposed a novel method for reconstructing pulse
signals that recur in time. The method is based on the study of the probability distribution of sample trains along curve \curve. Numerical simulations have shown that the
root-mean-square error of the proposed reconstruction drops with the number of pulse
recurrences $N$ at the rate $N^{-\frac{1}{2}}$. The proposed algorithm can
withstand small disturbances of the input data. However, it would require some
modifications at the stages of curve \curve and quantile function~\Qpdtau
estimation to make it more robust to noisy samples. These
modifications are expected to be similar to those proposed
in~\cite{Rup21icassp} in the context of periodic signal reconstruction.
Besides such adaptation, we work on extending the procedure to pulse streams consisting of pulses that vary in amplitude.

%
%
%



\bibliographystyle{IEEEtran}
\bibliography{IEEEabrv,biblio}

\begin{thebibliography}{10}
\providecommand{\url}[1]{#1}
\csname url@samestyle\endcsname
\providecommand{\newblock}{\relax}
\providecommand{\bibinfo}[2]{#2}
\providecommand{\BIBentrySTDinterwordspacing}{\spaceskip=0pt\relax}
\providecommand{\BIBentryALTinterwordstretchfactor}{4}
\providecommand{\BIBentryALTinterwordspacing}{\spaceskip=\fontdimen2\font plus
\BIBentryALTinterwordstretchfactor\fontdimen3\font minus \fontdimen4\font\relax}
\providecommand{\BIBforeignlanguage}[2]{{%
\expandafter\ifx\csname l@#1\endcsname\relax
\typeout{** WARNING: IEEEtran.bst: No hyphenation pattern has been}%
\typeout{** loaded for the language `#1'. Using the pattern for}%
\typeout{** the default language instead.}%
\else
\language=\csname l@#1\endcsname
\fi
#2}}
\providecommand{\BIBdecl}{\relax}
\BIBdecl

\bibitem{Rudresh18}
S.~Rudresh, S.~Nagesh, and C.~S. Seelamantula, ``Asymmetric pulse modeling for {FRI} sampling,'' \emph{IEEE Transactions on Signal Processing}, vol.~66, no.~8, pp. 2027--2040, 2018.

\bibitem{Pan17}
H.~Pan, T.~Blu, and M.~Vetterli, ``Towards generalized {FRI} sampling with an application to source resolution in radioastronomy,'' \emph{IEEE Transactions on Signal Processing}, vol.~65, no.~4, pp. 821--835, 2017.

\bibitem{Tur11}
R.~Tur, Y.~C. Eldar, and Z.~Friedman, ``Innovation rate sampling of pulse streams with application to ultrasound imaging,'' \emph{IEEE Transactions on Signal Processing}, vol.~59, no.~4, pp. 1827--1842, 2011.

\bibitem{Saurav20}
S.~K. Shastri, S.~Rudresh, R.~Anand, S.~Nagesh, C.~S. Seelamantula, and A.~K. Thittai, ``Axial super-resolution in ultrasound imaging with application to non-destructive evaluation,'' \emph{Ultrasonics}, vol. 108, p. 106183, 2020.

\bibitem{Matusiak12}
E.~Matusiak and Y.~C. Eldar, ``Sub-{N}yquist sampling of short pulses,'' \emph{IEEE Transactions on Signal Processing}, vol.~60, no.~3, pp. 1134--1148, 2012.

\bibitem{Hedge11}
C.~Hegde and R.~G. Baraniuk, ``Sampling and recovery of pulse streams,'' \emph{IEEE Transactions on Signal Processing}, vol.~59, no.~4, pp. 1505--1517, 2011.

\bibitem{Meisam20}
M.~Najjarzadeh and H.~Sadjedi, ``Implementation of particle swarm optimization algorithm for estimating the innovative parameters of a spike sequence from noisy samples via maximum likelihood method,'' \emph{Digital Signal Processing}, vol. 106, p. 102799, 2020.

\bibitem{Huang22}
G.~Huang, S.~Zhang, L.~Chen, H.~Han, and W.~Lu, ``Sub-{N}yquist sampling system for pulse streams based on non-ideal filters,'' \emph{Digital Signal Processing}, vol. 123, p. 103380, 2022.

\bibitem{Rup20icassp}
M.~W. Rupniewski, ``Triggerless random interleaved sampling,'' in \emph{{ICASSP} 2020 - 2020 {IEEE} International Conference on Acoustics, Speech and Signal Processing ({ICASSP})}, 2020, pp. 5605--5609.

\bibitem{Rup21icassp}
------, ``Super-resolution of periodic signals from short sequences of samples,'' in \emph{{ICASSP} 2021 - 2021 {IEEE} International Conference on Acoustics, Speech and Signal Processing ({ICASSP})}, 2021.

\bibitem{Rup21spl}
------, ``Reconstruction of periodic signals from asynchronous trains of samples,'' \emph{{IEEE} Signal Processing Letters}, vol.~28, pp. 289--293, 2021.

\bibitem{Rup21iet}
------, ``Period and signal reconstruction from the curve of trains of samples,'' \emph{{IET} Signal Processing}, vol.~16, no.~2, pp. 232--237, nov 2021.

\bibitem{dey2000}
T.~Dey, K.~Mehlhorn, and E.~Ramos, ``Curve reconstruction: Connecting dots with good reason,'' \emph{Computational Geometry: Theory and Applications}, vol.~15, no.~4, pp. 229--244, 2000.

\bibitem{lenz05}
T.~Lenz, ``Simple reconstruction of non-simple curves,'' Freie Universitat Berlin, Tech. Rep. B 05-02, 2005.

\end{thebibliography}

\end{document}